\documentclass[aps,prl,twocolumn,superscriptaddress]{revtex4-2}

\usepackage{amsmath,amssymb}
\usepackage{graphicx}
\usepackage{tikz}
\usetikzlibrary{arrows.meta, decorations.pathmorphing, calc}

\begin{document}  

\title{A Non-Markovian Route to Coherence in Heterogeneous Diffusive Systems}

\author{Aranyak Sarkar}

\affiliation{Experimental Physics I, University of Bayreuth, Universitätsstr. 30, 95447 Bayreuth, Germany}
\email{Aranyak.Sarkar@uni-bayreuth.de, aranyak@barc.gov.in}
\affiliation{Radiation \& Photochemistry Division, Bhabha Atomic Research Centre, Mumbai 400085, India}

\begin{abstract}
Temporal coherence—persistent alignment across time—can arise between agents with fundamentally distinct dynamics, a behavior that classical diffusion models (e.g., Brownian motion, fractional Brownian motion, generalized Langevin equation) are inherently limited in capturing, particularly under strong heterogeneity. We introduce the Coupled Memory Graph Process (CMGP), where dynamic interplay between internal memory and directed coupling enables synchronized behavior even in the absence of reciprocity. An active particle with long-range memory remains temporally coherent with a subdiffusive partner, despite mismatched scaling laws and asymmetric information flow. Bayesian optimization identifies a broad parameter regime supporting this phenomenon, characterized by a high State Persistence Index (\(\mathcal{SPI}\)). These results uncover a minimal mechanism for emergent coordination—a form of \emph{ghost coherence}—that remains inaccessible to classical stochastic models, with implications for viscoelastic environments and heterogeneous active systems.

\end{abstract}

\maketitle  

Stochastic dynamics underpin a wide range of processes in physics, biology, and soft matter—from intracellular organelle transport to stress relaxation in polymer networks and fluctuating flows in active media. Classical models such as Brownian motion (BM) and the Langevin equation have long served as foundational tools for describing such random motion. Yet experimental observations across diverse systems reveal significant deviations from purely Markovian behavior, including memory effects, heterogeneous diffusivity, and environment-mediated feedback \cite{Ratushny2012,PhysRevResearch2012057,weiss2004anomalous,metzler2000random, barkai2012strange, sokolov2012models, fakhri2014longrange,PhysRevE101032408,qiu2024nonmarkovian}. These findings challenge models that assume time-local, homogeneous noise and call for frameworks that can capture temporal persistence, cross-trajectory influence, and emergent correlations in non-equilibrium media.

To capture temporal correlations, models such as fractional Brownian motion (FBM) impose power-law scaling in the displacement autocorrelation, typically written as $\langle x(t)x(t+\tau) \rangle \sim \tau^{2H}$, where $H$ is the Hurst exponent \cite{mandelbrot1968fractional, jeon2011vivo}. The Generalized Langevin Equation (GLE) introduces memory via convolution with a friction kernel $\Gamma(t)$, yielding the integro-differential form
\begin{equation}
m\ddot{x}(t) + \int_0^t \Gamma(t-s)\dot{x}(s)\,ds = \xi(t)
\end{equation}
where $\xi(t)$ denotes correlated thermal noise \cite{kou2004generalized, lutz2001fractional}. While these models capture long-time correlations, they remain fundamentally trajectory-local—each particle evolves independently, without inter-trajectory feedback.

Other non-Markovian frameworks—such as aging fractional dynamics, power-law viscoelastic GLEs, and generalized continuous-time random walks—have extended memory descriptions \cite{barkai2003aging,jeon2013noisy, sokolov2012models,kou2004generalized}, yet still lack a mechanism for directional coupling between distinct entities.

This limitation becomes critical in structured, dynamic environments. In active biological media, for instance, motor-driven cargo deforms the viscoelastic matrix, which in turn modulates neighboring particles through stress propagation \cite{mizuno2007nonequilibrium}. These interactions are inherently asymmetric and time-lagged: the conditional expectation $\langle \Delta x_i(t)\Delta x_j(t+\tau) \rangle$ can be significant even when the reverse is negligible, violating temporal reciprocity~\citep{onsager1931reciprocal} and challenging classical memory-based models.

Furthermore, biological and soft-matter systems exhibit heterogeneous dynamics: subdiffusion, active bursts, and intermittent trapping may coexist within the same domain \cite{Hofling2013,caspi2000enhanced, golding2006physical, weber2012nonthermal}. A physically meaningful model must therefore allow for particles with distinct noise amplitudes $\sigma_i$, memory kernels $\Theta_i(\tau)$, and diffusivity exponents $\alpha_i$—yet remain capable of generating persistent cross-correlations. Such interactions are captured by structured coherence functions of the form:
\begin{equation}
\phi_{ij}(\tau) = \langle \Delta x_i(t)\Delta x_j(t+\tau) \rangle - \langle \Delta x_i(t) \rangle \langle \Delta x_j(t+\tau) \rangle
\end{equation}

While composite-kernel GLEs and aging models partially address temporal heterogeneity \cite{barkai2003aging,jeon2013noisy}, they treat memory as an intrinsic, symmetric property of isolated trajectories. Consequently, they fail to account for emergent phenomena such as asymmetric feedback, spatiotemporal coherence, and environmentally mediated cross-trajectory influence \cite{dieterich2008anomalous,lampo2017cytoplasmic,masson2014inferring}.
 Mechanically coupled GLEs \cite{goychuk2012viscoelastic,kupferman2004fractional, lutz2001fractional} (MechGLE) can generate alignment through shared viscoelastic baths, yet typically assume symmetric spring-like couplings and require similar diffusive states for coherence to persist. These constraints reflect Onsager reciprocity\cite{onsager1931reciprocal}, which enforces time-reversal symmetry and equal cross-response coefficients near equilibrium. Systems that exhibit non-reciprocal coherence—such as active cytoplasmic transport—thus demand stochastic models that permits asymmetry in both memory and interactions \cite{mizuno2007nonequilibrium,fakhri2014longrange}.

We introduce the \textit{Coupled Memory Graph Process} (CMGP), a unified framework that embeds heterogeneous memory and asymmetric temporal coupling to generate coherence between diffusively dissimilar particles. Formally, the evolution of each agent \( x_i(t) \) is governed by a non-Markovian integro-differential equation:

\begin{align}
\frac{dx_i(t)}{dt} &= \int_0^t \Theta_i(t{-}s)\, \xi_i(s)\, ds \notag\\
&\quad + \sum_{j \neq i} \int_0^t \Phi_{ij}(t,s)\, K_{ij}(d_{ij}(s))\, ds+ \eta_i(t)
\label{eq:cmgp_integral}
\end{align}

where \( \Theta_i(t) \) is the self-memory kernel shaping intrinsic temporal correlations, \( \Phi_{ij}(t,s) \) encodes directed temporal coherence from agent \( j \) to agent \( i \), and \( K_{ij}(d) \) is a distance-dependent gain function modulating interactions based on the instantaneous separation \( d_{ij}(s) = \| x_i(s) - x_j(s) \| \). The term \( \xi_i(s) \) represents colored or white noise driving the memory-based dynamics, while \( \eta_i(t) \) denotes an independent free noise term capturing direct, uncorrelated stochastic fluctuations.

This formalism maps directly onto Fig.~1, where each node \( x_i(t) \) integrates its own memory (via \( \Theta_i \), shown as blue self-loops) and receives delayed, distance-weighted input from others (via \( \Phi_{ij} \cdot K_{ij} \), red arrows). The coupling is inherently asymmetric—since \( \Phi_{ij} \ne \Phi_{ji} \)—enabling directional coordination even across mismatched diffusivities.

By intentionally violating Onsager reciprocity\citep{onsager1931reciprocal}, CMGP captures non-equilibrium dynamics inaccessible to classical models. Beyond pairwise interactions, it extends naturally to structured environments—such as deformable boundaries or viscoelastic fields—where memory-driven coupling with the medium produces long-range coherence. In such contexts, CMGP offers a flexible platform for modeling emergent, feedback-sensitive transport in soft and active matter.

\begin{figure}[h!]
\centering
\begin{tikzpicture}[scale=0.7, every node/.style={font=\scriptsize}, >=Stealth]

\node[circle, draw, thick, fill=blue!10] (x1) at (0,0) {$x_1(t)$};
\node[circle, draw, thick, fill=blue!10] (x2) at (3.5,0) {$x_2(t)$};

\draw[->, thick, blue] (x1) edge[out=135, in=225, looseness=6] node[left] {$\Theta_1(t)$} (x1);
\draw[->, thick, blue] (x2) edge[out=45, in=-45, looseness=6] node[right] {$\Theta_2(t)$} (x2);

\draw[->, thick, red] (x1) to[bend left=20] node[above] {\footnotesize $\Phi_{12}(t,\tau)\cdot K_{12}(t)$} (x2);
\draw[->, thick, red, dashed] (x2) to[bend left=20] node[below] {\footnotesize $\Phi_{21}(t,\tau)\cdot K_{21}(t)$} (x1);

\node[circle, draw=gray, thick, dotted, minimum size=18pt] (Xd) at (6.6, 0.0) {};
\node[gray] at (6.6,-0.57) {\scriptsize $x_3(t)$};

\end{tikzpicture}
\caption{\small 
Schematic representation of the Coupled Memory Graph Process (CMGP). Each particle $x_i(t)$ (representing A and B) integrates internal noise through a self-memory kernel $\Theta_i(t)$ (blue loops), while also receiving temporally lagged influence from others via $\Phi_{ij}(t,\tau)$. These coherence interactions are weighted by adaptive gains $K_{ij}(t)$, which may incorporate inter-particle distance or geometric constraints. The coupling is inherently asymmetric and dynamic. Dotted nodes (e.g., $x_3(t)$) illustrate extensibility to larger systems with heterogeneous connectivity and memory structure.
}
\label{fig:cmgp_schematic}
\end{figure}
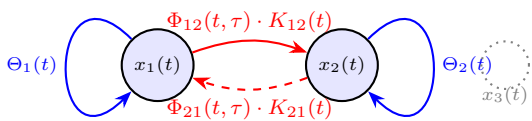

To understand how coherence emerges in complex networks of interacting agents, it is instructive to begin with the simplest nontrivial case: a two-particle system. This minimal setup captures the core mechanism of directionally coupled, memory-driven interactions while remaining analytically and computationally tractable.

We demonstrate the CMGP framework using a confined two-particle system in two-dimensional space, where particles \( A \) and \( B \) exhibit distinct diffusive behaviors. Particle \( A \) is \textit{active}, influenced both by its own memory and by input from particle \( B \). In contrast, particle \( B \) is \textit{passive}, evolving solely under its intrinsic memory without external input. This setup directly corresponds to the schematic shown in Fig.~1: each particle integrates its own past through a self-memory kernel \( \Theta(\tau) \) (illustrated by blue loops). Additionally, particle \( A \) receives history-dependent, distance-modulated input from particle \( B \) via a coherence kernel \( \Phi(\tau) \) weighted by the spatial gain function \( K(d_{AB}(t)) \) (represented by a red arrow).

Unlike classical models where symmetry or feedback ensures convergence, the interaction here is explicitly asymmetric: particle \( B \) influences \( A \), but not vice versa. This configuration enables us to test the central premise of CMGP: can directional, memory-mediated coupling induce persistent temporal coherence in the presence of vast diffusive heterogeneity?

Inspired by the Volterra framework\citep{volterra} for systems with memory, we formulate the discrete-time CMGP update rules as:
\begin{align}
x_A(t{+}1) &= x_A(t) + \sum_{\tau=1}^M \Theta(\tau)\, \xi_A(t{-}\tau) \notag\\
           &\quad + K(d_{AB}(t)) \sum_{\tau=1}^M \Phi(\tau)\, \Delta x_B(t{-}\tau) + \eta_A(t), \\
x_B(t{+}1) &= x_B(t) + \sum_{\tau=1}^M \Theta(\tau)\, \xi_B(t{-}\tau) + \eta_B(t),
\end{align}
where \( \xi_i(t) \) are temporally correlated noise inputs, \( \eta_i(t) \) are uncorrelated Gaussian fluctuations, and \( d_{AB}(t) = \|x_A(t) - x_B(t)\| \) denotes the instantaneous distance between particles. 

The CMGP model is governed by two key parameters: the \emph{coupling strength} \( \kappa \) and the \emph{memory depth} \( M \). The coupling strength \( \kappa \) controls the magnitude of inter-agent coherence, while the memory depth \( M \) defines how much of its own past each agent integrates over time. Spatial interactions between agents are modulated through a gain function \( K(d) = \kappa \exp(-d/R) \), where \( R \) is the confinement radius, which sets the characteristic decay range with inter-particle distance \( d(t) \). This exponential form captures biologically relevant attenuation mechanisms—such as mechanical stress propagation and diffusive signaling—where influence diminishes sharply with separation~\cite{Bauerle2018,trepat2009physical,edwards2011force}.

Temporal memory is modeled using power-law decaying kernels for both self-memory and cross-memory: \(\Theta(\tau) = (\tau + 1)^{-\alpha_m}\) and \(\Phi(\tau) = (\tau + 1)^{-\gamma_c}\), where \(\alpha_m, \gamma_c \in (0,1)\) control the decay of self- and coherence memory, respectively. These forms reflect long-range temporal correlations observed in subdiffusive, viscoelastic, and active systems.

The core idea of the CMGP framework is to introduce directionally biased, memory-driven interactions---unlike classical models such as Brownian motion (BM), fractional Brownian motion (FBM), or generalized Langevin equations (GLE), which either lack inter-particle coupling or rely on symmetric feedback (as in MechGLE). These asymmetric pathways enable particles to remain temporally synchronized without requiring convergence in their diffusive behaviors. To quantitatively assess this behavior, we define three core descriptors. The time-averaged mean squared displacement (TA-MSD) is given by \(\langle \delta^2_i(\tau) \rangle = \langle (x_i(t{+}\tau) - x_i(t))^2 \rangle_t\), capturing the typical displacement of particle \(i\) over lag \(\tau\). The local diffusion exponent \(\alpha_i(\tau) = \frac{d \log_{10} \langle \delta^2_i(\tau) \rangle}{d \log_{10} \tau}\) provides a lag-resolved scaling profile; while transient variations in \(\alpha_i(\tau)\) reflect time-local dynamical behavior, its asymptotic value at large lags, \(\tau \to \infty\), approaches the effective Hurst exponent, \(\alpha_i(\tau) \to 2H\), thus defining the true diffusive state. Temporal coherence is quantified by the normalized cross-correlation \(C_{AB}(\tau) = \frac{ \langle \Delta x_A(t, \tau)\, \Delta x_B(t, \tau) \rangle_t }{ \sqrt{ \langle \Delta x_A^2(t, \tau) \rangle_t \langle \Delta x_B^2(t, \tau) \rangle_t } }\), where \(\Delta x_i(t, \tau) = x_i(t{+}\tau) - x_i(t)\). The CMGP model uniquely supports the coexistence of high temporal coherence and persistent diffusive disparity. This regime---marked by elevated \(C_{AB}(\tau)\) and sustained differences in \(\alpha_i(\tau)\)---defines the emergent phenomenon of \emph{ghost coherence}, where particles are ‘in sync’ in time without mirroring each other’s trajectories, a feature absent in conventional stochastic models, which invariably require convergence of dynamics to achieve long-term correlations. Note that all variables in this study are expressed in arbitrary units to emphasize universal dynamical features, independent of specific material properties.

\begin{figure}[t]
    \centering
    \includegraphics[width=\columnwidth]{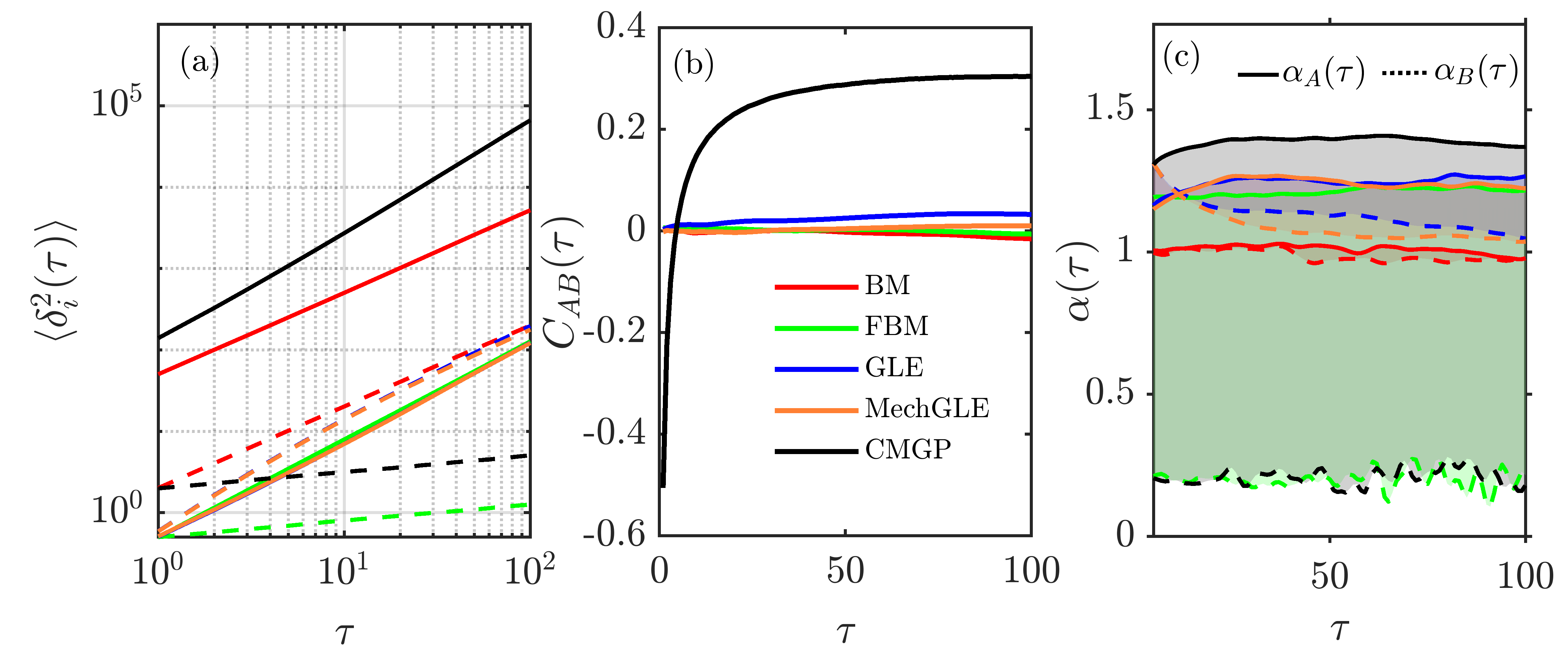}
    \caption{%
Quantitative comparison of two-particle dynamics under five models: Brownian motion (BM), fractional Brownian motion (FBM), generalized Langevin equation (GLE) with shared noise, symmetrically coupled MechGLE, and asymmetrically coupled CMGP. 
(a) Time-averaged MSD curves for particles \( A \) (solid) and \( B \) (dashed) show strong convergence in MechGLE and GLE (with shared noise), but persistent disparity in CMGP. 
(b) Temporal coherence \( C_{AB}(\tau)\) is negligible in BM, FBM, moderate in GLE,,MechGLE, and sustained in CMGP. 
(c) Local diffusion exponents \( \alpha_A(\tau) \) (solid) and \( \alpha_B(\tau) \) (dashed) illustrate loss of heterogeneity in MechGLE/GLE, while CMGP preserves it—demonstrating the hallmark of ghost coherence.
}
    \label{fig:cmgp_comparison}
\end{figure}

 Fig.~\ref{fig:cmgp_comparison} highlights the distinct dynamical signatures of classical and memory-driven models. In BM, FBM, and uncoupled GLE, particles \( A \) and \( B \) follow uncorrelated, isotropic motion, with MSDs that remain clearly separated across all lag times (Fig.~\ref{fig:cmgp_comparison}a). However, in our GLE implementation, we introduce a shared colored noise source to mimic a common viscoelastic environment. This shared stochastic background induces mild temporal alignment between particles, causing their trajectories and diffusion exponents to drift closer over time—even in the absence of explicit coupling. This results in a small but nonzero coherence \( C_{AB}(\tau) \) and narrowing of the gap between \( \alpha_A(\tau) \) and \( \alpha_B(\tau) \) (Fig.~\ref{fig:cmgp_comparison}b,c).

In MechGLE, we initialize particles \(A\) and \(B\) with distinct memory exponents—\(\gamma_A\) and \(\gamma_B\)—which govern the power-law decay of their respective memory kernels (e.g., \(\gamma_A = 0.8\) and \(\gamma_B = 1.9\)), to examine how quickly mechanical coupling drives dynamical convergence. Owing to the symmetric nature of their coupling through a mechanical spring, this convergence occurs rapidly, with both particles soon exhibiting identical effective dynamics—a hallmark of conservative, symmetry-enforcing interactions. Although high temporal coherence can be achieved in both GLE and MechGLE by suitable parameter tuning, it invariably comes at the cost of losing diffusive heterogeneity. This intrinsic trade-off, rooted in the reciprocal coupling structure, highlights a fundamental limitation of these models in sustaining coherent yet distinct dynamical behaviors.

By contrast, the CMGP framework preserves dynamical disparity while maintaining sustained coherence. Despite persistent differences between \( \alpha_A(\tau) \) and \( \alpha_B(\tau) \), the cross-correlation \( C_{AB}(\tau) \) remains high. This coexistence of coherence without convergence defines the regime of \emph{ghost coherence}, which emerges uniquely through asymmetric, memory-mediated interactions~\cite{volterra,kubo1966}---a hallmark of CMGP dynamics that closely parallels the ``anomalous yet Brownian'' diffusion scenario~\cite{PhysRevLett113098302}. Asymmetric coupling phenomena, such as leader-follower synchronization~\cite{Jadbabaie2003} in networks or directed percolation~\cite{Hinrichsen2000} in statistical physics, capture aspects of one-way influence but typically lead to full dynamical convergence or static transitions. In contrast, the CMGP model uniquely supports high temporal coherence while preserving persistent dynamical disparity. Notably, CMGP coherence curves exhibit an initial negative dip (e.g., \(C_{AB}(1) \approx -0.5\)), reflecting a structural phase mismatch: at early times (\(\tau \approx 0\)), particle \(A\) has no motion of \(B\) to follow, leading to transient anti-correlation. As \(B\) begins to move, \(A\) dynamically mimics its behavior with a delay, and coherence builds up over time through memory accumulation. Thus, \emph{ghost coherence} here is fundamentally mimetic rather than genetic—an emergent alignment shaped by ongoing dynamics, not preset symmetry.

\begin{figure}[t]
    \centering
    \includegraphics[width=\columnwidth]{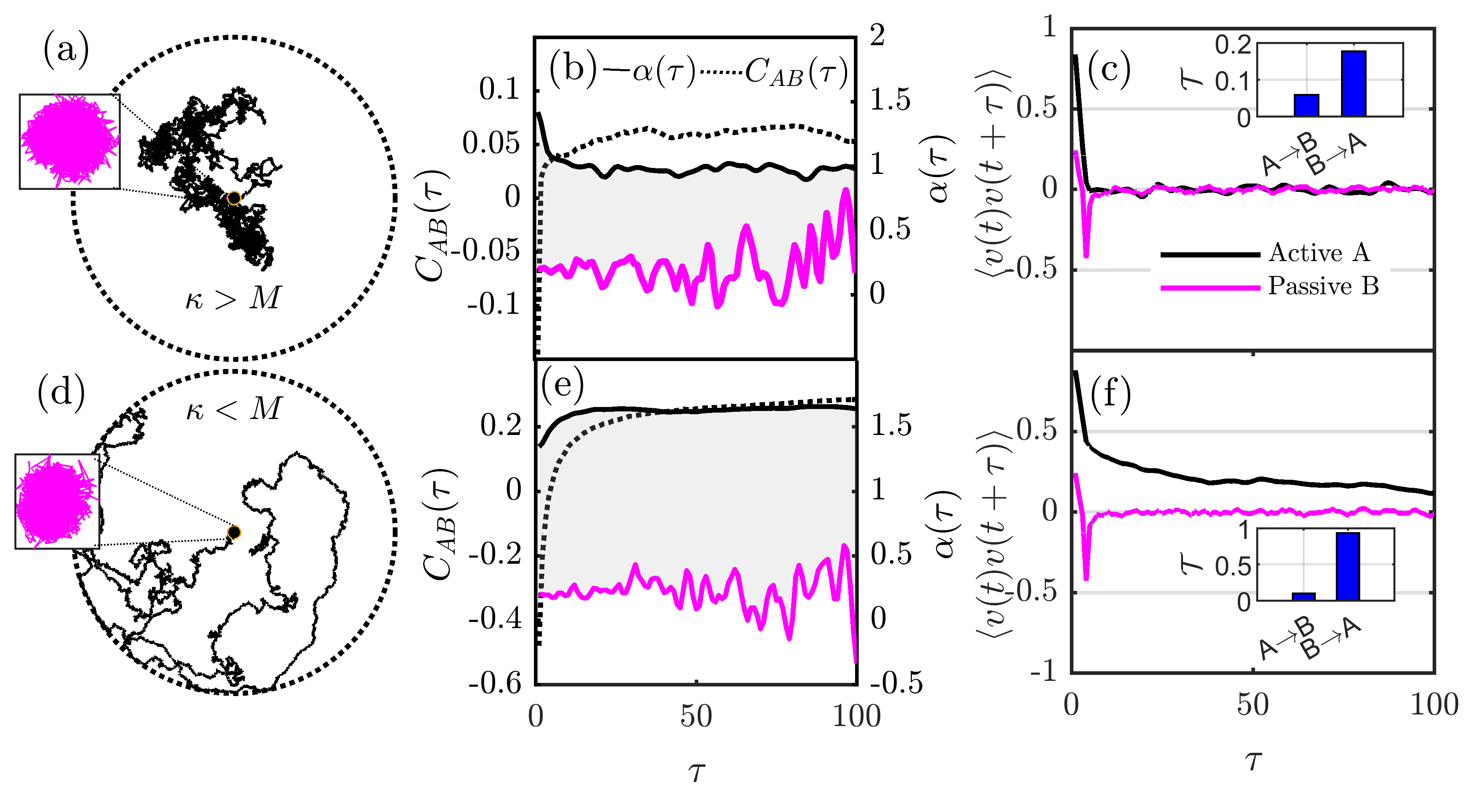}
    \caption{%
Representative CMGP realizations illustrating how the balance between coherence strength \( \kappa \) and memory depth \( M \) shapes trajectory structure and inter-particle dynamics. 
(a,d) Trajectories of active particle \( A \) (black) and passive particle \( B \) (magenta); inset: zoom on passive motion. 
Top row: \( \kappa > M (\kappa/M=10, R=150)\) regime; bottom row: \( \kappa < M (\kappa/M=0.1, R=600)\) regime. 
(b,e) Cross-coherence \( C_{AB}(\tau) \) and local diffusion exponents \( \alpha(\tau) \) reveal persistent directional alignment in both cases, with greater disparity in \( \alpha \) when memory dominates. 
(c,f) Velocity autocorrelation functions \( \langle v(t)v(t{+}\tau) \rangle \) and transfer entropy~\cite{schreiber2000} bar plots confirm consistent unidirectional information flow from the passive to active particle, despite markedly different spatial and temporal signatures. 
Together, these results demonstrate that CMGP supports structurally distinct but coherently coupled dynamics, tunable via the interplay between memory and coupling strength.
}
    \label{fig:cmgp_kappa_memory}
\end{figure}

Fig.~\ref{fig:cmgp_kappa_memory} compares two realizations of the CMGP model that demonstrate persistent directional coherence despite contrasting spatial trajectories, driven by different balances between coupling strength ($\kappa$) and memory depth ($M$). In the top row ($\kappa > M$), the active particle A maintains a localized and temporally consistent trajectory, whereas in the bottom row ($\kappa < M$), A exhibits broader spatial wandering. In both cases, the passive particle B remains confined, and the overall cross-correlation $C_{AB}(\tau)$ (panels b,e) remains high—indicating that coherent alignment persists independent of visual or geometric similarity. Panels (c,f) present the velocity autocorrelation functions (VACF), $\langle v(t)v(t{+}\tau) \rangle$, which reveal long-term temporal structure in A’s motion, particularly when its memory dominates (i.e., large $M$). Transfer entropy~\cite{schreiber2000}, defined as the excess uncertainty reduction in predicting the future of particle \(j\) given the past of particle \(i\) beyond the past of \(j\) itself, quantifies the directed information flow between coupled dynamics. Mathematically, it is given by \(\mathcal{T}_{i \rightarrow j} = \sum p(x_j^{t+1}, x_j^t, x_i^t) \log \left( \frac{p(x_j^{t+1} \mid x_j^t, x_i^t)}{p(x_j^{t+1} \mid x_j^t)} \right)\), where \(p(\cdot)\) denotes joint probabilities. Despite symmetric mechanical coupling, the observed transfer entropy \(\mathcal{T}_{i \rightarrow j}\) (insets) consistently exhibits a pronounced asymmetry: \(\mathcal{T}_{B \rightarrow A} \gg \mathcal{T}_{A \rightarrow B}\), confirming that the effective information flow is predominantly unidirectional. Notably, this asymmetry is amplified when particle \(A\) preserves its internal persistence—when memory dominates, \(A\) maintains its own dynamic integrity while becoming increasingly susceptible to informative influence from \(B\), despite not directly mirroring its motion.

The CMGP dynamics resemble a kite loosely tethered to a jittery pillar: although the kite appears free, its coherent motion subtly encodes the pillar’s fluctuations. Attempts by the kite to control the anchor disrupt coherence, whereas passive adherence enables it to act as an ideal information carrier, persistently encoding the partner’s dynamics. This highlights how asymmetry fosters functional coherence: the more persistently \(A\) preserves its own behavior, the more effectively it transmits information from \(B\). Thus, CMGP reveals a non-reciprocal coordination mechanism, where persistent self-dynamics coexist with directional informational guidance, challenging classical notions of mutual interaction.

\begin{figure}[t]
    \centering
    \includegraphics[width=0.71\columnwidth]{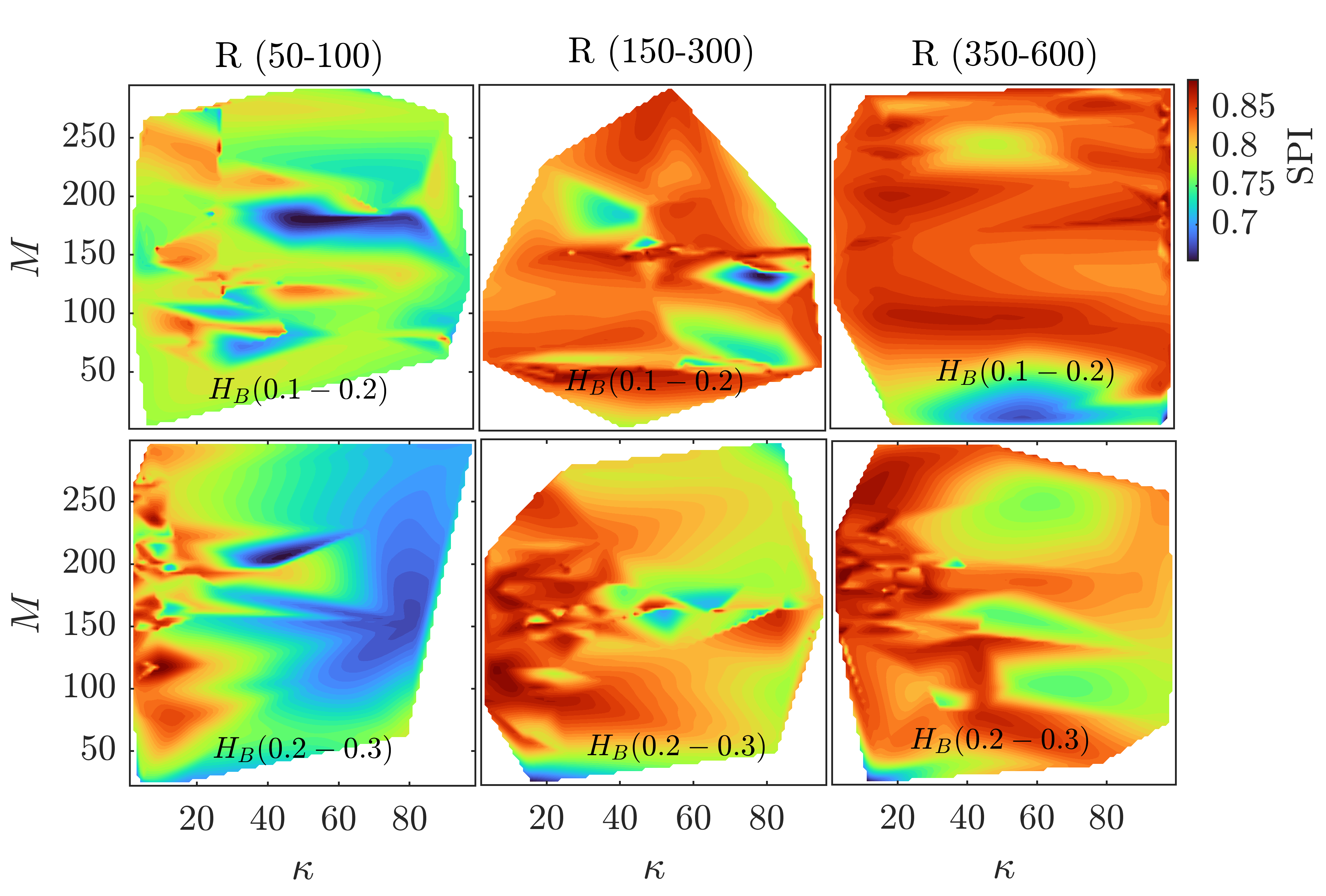}
    \caption{
    Phase diagrams of the State Persistence Index ($\mathcal{SPI}$) across the CMGP parameter space:
    $\mathcal{SPI}$ is defined as $1 - std(\alpha_A)$, quantifying the temporal consistency of the active particle's local scaling exponent.
    Each panel displays $\mathcal{SPI}$ as a function of coupling strength $\kappa$ and memory depth $M$, across varying ranges of confinement radius $R$ (columns) and passive Hurst exponent $H_B$ (rows).
    Warmer colors indicate more stable (persistent) motion of the active agent, while cooler regions reflect dynamic instability.
    The results demonstrate that sufficient mechanical space ($R$) enables A to sustain coherent dynamics while maintaining responsiveness to the subdiffusive background B.
    }
    \label{fig:spi_column}
\end{figure}
To identify regimes where coherence and persistence coexist, we employed Bayesian optimization~\cite{mackay2003} to maximize the State Persistence Index ($\mathcal{SPI}$), defined as \(\mathcal{SPI} = 1 - \mathrm{std}(\alpha_A)\), over the CMGP parameter space \((\kappa, M, R, H_B)\). Treating the $\mathcal{SPI}$ landscape as a posterior distribution \(p(\mathcal{SPI} \mid \kappa, M, R, H_B)\), new evaluations were selected by optimizing an acquisition function balancing exploration (uncertainty) and exploitation (predicted gain). Fig.~\ref{fig:spi_column} shows Bayesian-smoothed phase maps of $\mathcal{SPI}$ over \((\kappa, M)\) for varying \((R, H_B)\). As \(R\) increases and \(H_B\) decreases (\(H_B \lesssim 0.3\)), broader stability zones emerge, confirming the "kite" intuition: spatial freedom enhances dynamic self-consistency while maintaining directional coherence. Thus, robust coherence persists across a wide parameter swath without fine-tuning, arising from constructive memory-coupling interplay under spacious confinement.

Building on these phase-mapping results, our findings establish that two particles with distinct diffusive identities can sustain persistent temporal coherence without requiring trajectory symmetry or mutual adaptation. This robustness—emerging naturally across wide regions of the CMGP parameter space—highlights a new paradigm where relational memory, rather than mechanical consensus, underpins coherence. Notably, in complex, heterogeneous, and disordered environments, visual intuition may no longer align with temporal synchrony, as \emph{ghost} coherence can be deeply encoded in memory-driven dynamics rather than spatial proximity. Echoing principles observed in intracellular transport, CMGP dynamics reveal that stability and information integration can coexist without sacrificing dynamical individuality. The asymmetry in transfer entropy further underscores that influence need not flow symmetrically to be functionally meaningful. Looking forward, CMGP opens a powerful new avenue for modeling heterogeneous, memory-rich networks, where coherence is not imposed but self-organized—a direction promising to deepen our understanding of complex active systems and inspire design principles for synthetic active materials.

\bibliographystyle{unsrt}     
\bibliography{new_refs}           

\begin{thebibliography}{10}

\bibitem{Ratushny2012}
Alexander~V. Ratushny, Rasha~A. Saleem, Kelly Sitko, Stephen~A. Ramsey, and
  Jennifer~D. Aitchison.
\newblock Asymmetric positive feedback loops reliably control biological
  responses.
\newblock {\em Molecular Systems Biology}, 8:577, 2012.

\bibitem{PhysRevResearch2012057}
N.~Levernier, O.~B\'enichou, R.~Voituriez, and T.~Gu\'erin.
\newblock Kinetics of rare events for non-markovian stationary processes and
  application to polymer dynamics.
\newblock {\em Phys. Rev. Res.}, 2:012057, Mar 2020.

\bibitem{weiss2004anomalous}
Matthias Weiss, Martin Elsner, Fredrik Kartberg, and Tommy Nilsson.
\newblock Anomalous subdiffusion is a measure for cytoplasmic crowding in
  living cells.
\newblock {\em Biophysical Journal}, 87(5):3518--3524, 2004.

\bibitem{metzler2000random}
Ralf Metzler and Joseph Klafter.
\newblock The random walk's guide to anomalous diffusion: a fractional dynamics
  approach.
\newblock {\em Physics Reports}, 339(1):1--77, 2000.

\bibitem{barkai2012strange}
Eli Barkai, Yuval Garini, and Ralf Metzler.
\newblock Strange kinetics of single molecules in living cells.
\newblock {\em Physics Today}, 65(8):29--35, 2012.

\bibitem{sokolov2012models}
Igor~M. Sokolov.
\newblock Models of anomalous diffusion in crowded environments.
\newblock {\em Soft Matter}, 8(35):9043--9052, 2012.

\bibitem{fakhri2014longrange}
Nikta Fakhri, Alexander~D. Wessel, Christian Willms, Matteo Pasquali, Dieter~R.
  Klopfenstein, Fred~C. MacKintosh, and Christoph~F. Schmidt.
\newblock Long-range correlated motion in the cytoplasm of living cells.
\newblock {\em Science}, 344(6187):1031--1035, 2014.

\bibitem{PhysRevE101032408}
Bernhard~G. Mitterwallner, Christoph Schreiber, Jan~O. Daldrop, Joachim~O.
  R\"adler, and Roland~R. Netz.
\newblock Non-markovian data-driven modeling of single-cell motility.
\newblock {\em Phys. Rev. E}, 101:032408, Mar 2020.

\bibitem{qiu2024nonmarkovian}
Yunrui Qiu, Rafal~P. Wiewiora, Jesus~A. Izaguirre, Huafeng Xu, Woody Sherman,
  Weiping Tang, and Xuhui Huang.
\newblock Non-markovian dynamic models identify non-canonical kras-vhl
  encounter complex conformations for novel protac design.
\newblock {\em JACS Au}, 4(10):3857--3868, 2024.

\bibitem{mandelbrot1968fractional}
Benoit~B. Mandelbrot and John~W. Van~Ness.
\newblock Fractional brownian motions, fractional noises and applications.
\newblock {\em SIAM Review}, 10(4):422--437, 1968.

\bibitem{jeon2011vivo}
Jae-Hyung Jeon, Vinicius Tejedor, Stanislav Burov, Eli Barkai,
  C.~Selhuber-Unkel, K.~Berg-S{\o}rensen, L.B. Oddershede, and Ralf Metzler.
\newblock In vivo anomalous diffusion and weak ergodicity breaking of lipid
  granules.
\newblock {\em Physical Review Letters}, 106(4):048103, 2011.

\bibitem{kou2004generalized}
S.C. Kou and X.S. Xie.
\newblock Generalized langevin equation with fractional gaussian noise:
  subdiffusion within a single protein molecule.
\newblock {\em Physical Review Letters}, 93(18):180603, 2004.

\bibitem{lutz2001fractional}
Eric Lutz.
\newblock Fractional langevin equation.
\newblock {\em Physical Review E}, 64(5):051106, 2001.

\bibitem{barkai2003aging}
Eli Barkai.
\newblock Aging in subdiffusion generated by a deterministic dynamical system.
\newblock {\em Physical Review Letters}, 90(10):104101, 2003.

\bibitem{jeon2013noisy}
Jae-Hyung Jeon and Ralf Metzler.
\newblock Noisy continuous time random walks.
\newblock {\em Journal of Physics A: Mathematical and Theoretical},
  46(34):345002, 2013.

\bibitem{mizuno2007nonequilibrium}
Daisuke Mizuno, C{\'e}cile Tardin, Christoph~F Schmidt, and Fred~C MacKintosh.
\newblock Nonequilibrium mechanics of active cytoskeletal networks.
\newblock {\em Science}, 315(5810):370--373, 2007.

\bibitem{onsager1931reciprocal}
Lars Onsager.
\newblock Reciprocal relations in irreversible processes. i.
\newblock {\em Physical Review}, 37(4):405--426, 1931.

\bibitem{Hofling2013}
Felix Höfling and Thomas Franosch.
\newblock Anomalous transport in the crowded world of biological cells.
\newblock {\em Reports on Progress in Physics}, 76(4):046602, 2013.

\bibitem{caspi2000enhanced}
Ayelet Caspi, Rony Granek, and Michael Elbaum.
\newblock Enhanced diffusion in active intracellular transport.
\newblock {\em Physical Review Letters}, 85(26):5655, 2000.

\bibitem{golding2006physical}
Ido Golding and Edward~C. Cox.
\newblock Physical nature of bacterial cytoplasm.
\newblock {\em Physical Review Letters}, 96(9):098102, 2006.

\bibitem{weber2012nonthermal}
Stephanie~C Weber, Andrew~J Spakowitz, and Julie~A Theriot.
\newblock Nonthermal atp-dependent fluctuations contribute to the in vivo
  motion of chromosomal loci.
\newblock {\em Proceedings of the National Academy of Sciences},
  109(19):7338--7343, 2012.

\bibitem{dieterich2008anomalous}
P.~Dieterich, R.~Klages, R.~Preuss, and A.~Schwab.
\newblock Anomalous dynamics of cell migration.
\newblock {\em Proceedings of the National Academy of Sciences},
  105(2):459--463, 2008.

\bibitem{lampo2017cytoplasmic}
Thomas~J. Lampo, Stella Stylianidou, Mikael~P. Backlund, Paul~A. Wiggins, and
  Andrew~J. Spakowitz.
\newblock Cytoplasmic rna-protein particles exhibit non-gaussian subdiffusive
  behavior.
\newblock {\em Biophysical Journal}, 112(3):532--542, 2017.

\bibitem{masson2014inferring}
Jean-Baptiste Masson, Pierre Dionne, Claudio Salvatico, Matthieu Renner,
  Christian~G. Specht, Antoine Triller, and Maxime Dahan.
\newblock Inferring maps of forces inside cell membranes from single-molecule
  trajectories.
\newblock {\em Nature Methods}, 11(5):571--573, 2014.

\bibitem{goychuk2012viscoelastic}
Igor Goychuk.
\newblock Viscoelastic subdiffusion: from anomalous to normal.
\newblock {\em Physical Review E}, 86(2):021113, 2012.

\bibitem{kupferman2004fractional}
Raz Kupferman.
\newblock Fractional kinetics in kac-zwanzig heat bath models.
\newblock {\em Journal of Statistical Physics}, 114(1-2):291--326, 2004.

\bibitem{volterra}
Vito Volterra.
\newblock {\em Theory of Functionals and of Integral and Integro-Differential
  Equations}.
\newblock Dover Publications, 1959.
\newblock Originally published in 1930.

\bibitem{Bauerle2018}
Tobias Bäuerle, Andreas Fischer, Thomas Speck, and Clemens Bechinger.
\newblock Self-organization of active particles by quorum sensing rules.
\newblock {\em Nature Communications}, 9:3232, 2018.

\bibitem{trepat2009physical}
Xavier Trepat, Matthew~R. Wasserman, Thomas~E. Angelini, Edward Millet,
  David~A. Weitz, James~P. Butler, and Jeffrey~J. Fredberg.
\newblock Physical forces during collective cell migration.
\newblock {\em Nature Physics}, 5(6):426--430, 2009.

\bibitem{edwards2011force}
Corey~M. Edwards and Ulrich~S. Schwarz.
\newblock Force localization in contracting cell layers.
\newblock {\em Physical Review Letters}, 107(12):128101, 2011.

\bibitem{kubo1966}
R.~Kubo.
\newblock The fluctuation-dissipation theorem.
\newblock {\em Reports on Progress in Physics}, 29(1):255--284, 1966.

\bibitem{PhysRevLett113098302}
Mykyta~V. Chubynsky and Gary~W. Slater.
\newblock Diffusing diffusivity: A model for anomalous, yet brownian,
  diffusion.
\newblock {\em Phys. Rev. Lett.}, 113:098302, Aug 2014.

\bibitem{Jadbabaie2003}
Ali Jadbabaie, Jie Lin, and A.~Stephen Morse.
\newblock Coordination of groups of mobile autonomous agents using nearest
  neighbor rules.
\newblock {\em IEEE Transactions on Automatic Control}, 48(6):988--1001, 2003.

\bibitem{Hinrichsen2000}
Haye Hinrichsen.
\newblock Non-equilibrium critical phenomena and phase transitions into
  absorbing states.
\newblock {\em Advances in Physics}, 49(7):815--958, 2000.

\bibitem{schreiber2000}
T.~Schreiber.
\newblock Measuring information transfer.
\newblock {\em Physical Review Letters}, 85(2):461--464, 2000.

\bibitem{mackay2003}
David J.~C. MacKay.
\newblock {\em Information Theory, Inference, and Learning Algorithms}.
\newblock Cambridge University Press, 2003.

\end{thebibliography}

\end{document}